\documentclass[11pt]{article}

\usepackage{palatino}

\usepackage{authblk}
\usepackage{graphicx}
\usepackage{hyperref}
\usepackage[margin=1.25in]{geometry}
\usepackage{xcolor}

\usepackage[numbers,sort&compress]{natbib}
\bibpunct{(}{)}{,}{n}{}{;}  

\begin{document}


\title{Perspective on Data Science}

\author[1]{Roger D. Peng}
\author[2]{Hilary S. Parker}
\affil[1]{Department of Biostatistics, Johns Hopkins Bloomberg School of Public Health}
\affil[2]{Independent Consultant, San Francisco, California}

\maketitle

\clearpage

\begin{abstract}
The field of data science currently enjoys a broad definition that includes a wide array of activities which borrow from many other established fields of study. Having such a vague characterization of a field in the early stages might be natural, but over time maintaining such a broad definition becomes unwieldy and impedes progress. In particular, the teaching of data science is hampered by the seeming need to cover many different points of interest. Data scientists must ultimately identify the core of the field by determining what makes the field unique and what it means to develop new knowledge in data science. In this review we attempt to distill some core ideas from data science by focusing on the iterative process of data analysis and develop some generalizations from past experience. Generalizations of this nature could form the basis of a theory of data science and would serve to unify and scale the teaching of data science to large audiences.
\end{abstract}

\clearpage

\section{Introduction}

Data science is perhaps one of the most generic and vaguely defined fields of study to have come about in the past 50 years. In a recent paper titled ``50 Years of Data Science'' David Donoho cites a number of definitions that ultimately could be interpreted to include essentially any scientific activity~\citep{Donoho2017}. To the extent that there is a common theme to the various definitions of data science, it is the decoupling of operational activity on \textit{data} from the \textit{scientific, policy, or business considerations} that may surround such activity. The key recent development is that this operational activity has grown signficantly more complex with the increase in dataset sizes and the growth in computational power~\citep{peng:2011}. It is therefore valuable to consider what distinguishes data science as a field of study and whether people who consider themselves members of the field share anything in common or even agree on the definition of the field.

Equal in difficulty to the task of defining the field of data science is defining what it is that data scientists do. On the U.C.~Berkeley School of Information's web site titled ``What is Data Science?" (https://ischoolonline.berkeley.edu/data-science/what-is-data-science/) they write, ``Data scientists examine which questions need answering and where to find the related data. They have business acumen and analytical skills as well as the ability to mine, clean, and present data. Businesses use data scientists to source, manage, and analyze large amounts of unstructured data. Results are then synthesized and communicated to key stakeholders to drive strategic decision-making in the organization." Such a definition is not uncommon in our experience and encompasses a wide range of possible activities that require a diverse set of skills from an individual. 

The common theme in descriptions of the job of the data scientist is a kind of ``beginning-to-end" narrative, whereby data scientists have a hand in many if not all aspects of a process that involves data. The only aspects in which they are not involved is the choice of the question itself and the decision that is ultimately made upon seeing the results. In fact, based on our experience, many real-world situations draw the data scientist into participating in those activites as well.

Having a vague and generic definition for a field of study, especially one so new in formation as data science, can provide security and other advantages in the short term. Drawing as many people as possible into a field can build strength and momentum that might subsequently lead to greater resources. It would seem premature to narrowly define a field in the early stages and risk excluding individuals that might make useful contributions. However, retaining such a vague definition of a field introduces challenges that ultimately limit progress. Over time, the ever-changing and quickly-advancing nature of the field leads to a greater number of activities and tools being included into the field, making it increasingly difficult to define what are the core elements of the field. A persistent temptation to define the field as simply the union of all activities by members of the field can lead to a kind of field-specific entropy.

The inclusion of a large number of activities into the definition of a field incurs little cost until one is confronted with teaching the field to newcomers~\citep{kross2019practitioners}. Students, or learners of any sort, arrive with limited time and resources and typically need to learn the ``essentials" or \textit{core} of the field. What exactly composes this core set of elements can be greatly influenced by the instructor's personal experience in the field of data science, which is unsurprising given the heterogeneity of people included in the field. The result is that different individuals can end up telling very different stories about what makes up the core of data science~\citep{wing:2020}. Computer scientists, statisticians, engineers, information scientists, and mathematicians~(to name but a few) will all focus on their specific perspective and relay their experience of what is important. Such a fracturing of the teaching of data science suggests that there is little about data science that is generalizable and that material should essentially be taught on a ``need to know" or ``just in time" basis. In that case, there is little rationale for a standardized formal education in data science.

It would be naive to think that the definition of a field ever becomes stable or that members of a field ever reach agreement on its definition. In fact, it is healthy for members of any field to question the fundamental core of the field itself and consider whether anything should be added or subtracted. In a 1962 paper in the \textit{Annals of Mathematical Statistics}, John Tukey debated whether data analysis was a part of statistics or a field unto its own~\citep{tuke:1962}. More recently, software engineering and computer science topics have been added to numerous statistics curricula as a fundamental aspect of the field~\citep{nolanlang2010}. No single paper or review will settle the matter of what makes the core of the field, but constant discussion and iteration has its own intellectual benefits and ensures that a field does not stagnate to the point of irrelevance.

At this point, it may be worthwhile to examine the long list of data science activities, tools, skills, and jobs requirements and determine if there exists any common elements---some things that all data scientists do or some tools that all data scientists use. The extent to which we can draw anything out of such an exercise will give us some indication of whether a core of data science exists and where its boundaries may lie.

\section{Building a Core of Data Science}

We will begin the discussion of the core of data science with data analysis. For most data scientists it is likely that data analysis plays some role in their routine activity and that tools for doing data analysis are commonly used. Whether a data analysis is the end product or an intermediate output on the path to a different end product will depend on the circumstances and setting. Nevertheless, it seems fair to say that at some point, a data scientist will analyze data.

Tukey defined data analysis as ``procedures for analyzing data, techniques for interpreting the results of such procedures, ways of planning the gathering of data to make its analysis easier, more precise or more accurate, and all the machinery and results of (mathematical) statistics which apply to analyzing data."~\citep{tuke:1962} If one were to develop a definition for today's world, one might make explicit mention of computing, workflow development, and the implementation of procedures on large-scale datasets. Data analysis itself has a bit of a hazy definition, with some characterizing it narrowly as the application of statistical methods to clean datasets and with others broadening its definition to the point where there is little distinction between data analysis and data science~\citep{wing:2020,chatfield1995,Donoho2017,wing:jane:kloe:2018}. In order to avoid getting caught in a never-ending recursion, we will assume for the moment that the definition of data analysis lies somewhere in the middle of that spectrum.

\subsection{A Data Analytic Triangle}

\begin{figure}[tbh]
\centering
\includegraphics[width=3in]{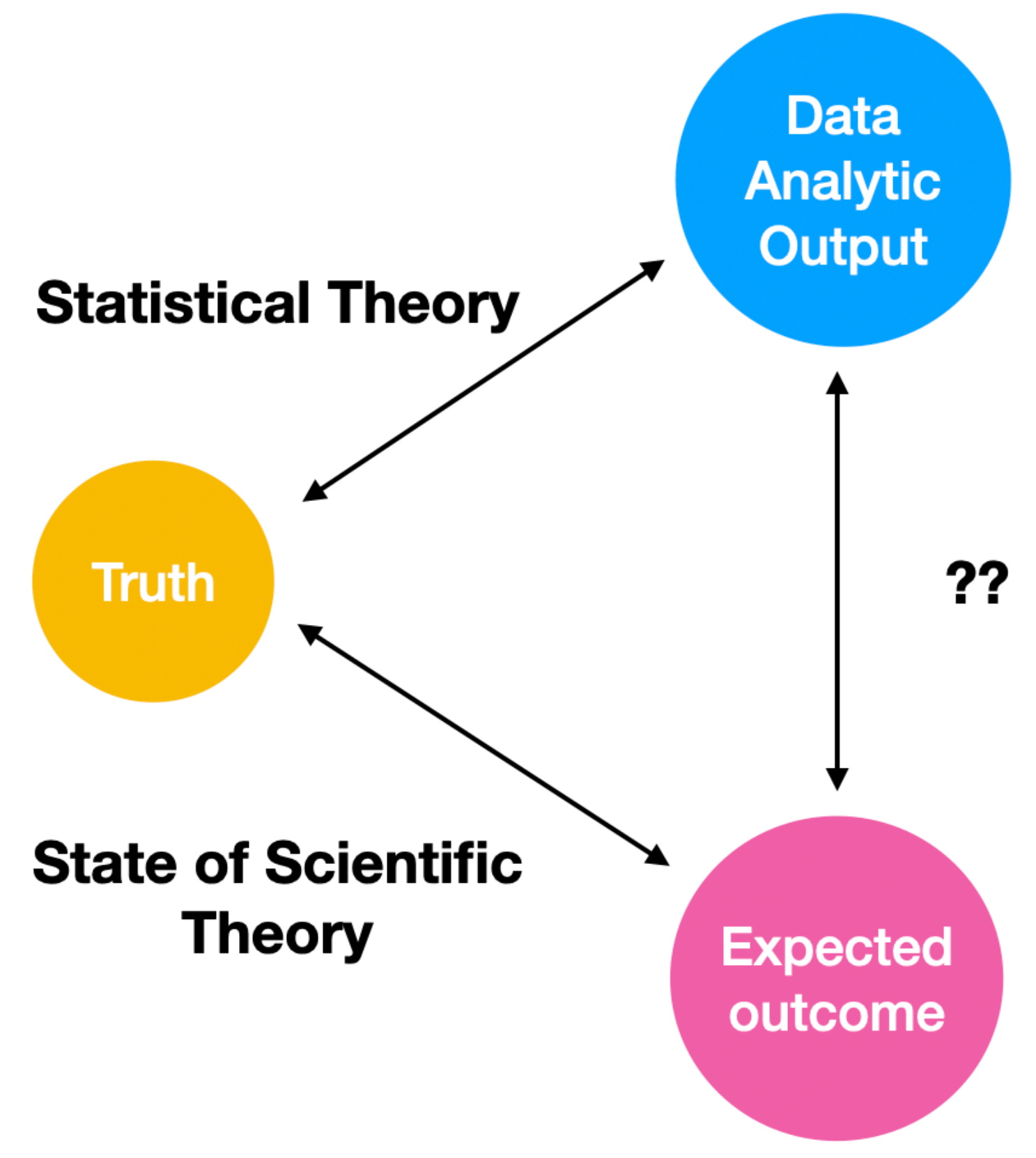}
\caption{``Triangle" of data analysis.}
\label{fig:triangle}
\end{figure}

Interaction with data at any stage of an analysis generally requires careful thought and consideration in order to make meaningful interpretations at the end. This idea that data scientists should ``carefully consider the data" can be made more explicit and will serve as the basis of the first part of our data science core. 
At any given instance of dealing with data we are engaged in a three part process in which the data scientist only has direct influence over two parts. The three parts are
\begin{enumerate}
\item \textit{Truth}: The underlying aspect of the world about which we are trying to learn;
\item \textit{Expected Outcome}: Our expectations for how a given data analytic output will be realized; and
\item \textit{Data Analytic Output}: The observed output from any data analytic action.
\end{enumerate}
These three parts can be arranged in a kind of ``triangle" as shown in Figure~\ref{fig:triangle}. 

The difference between the truth and one's data analytic output can typically be explained by statistical theory, which explicitly accounts for random deviations in the data. The difference between the truth and what one expects to see from an analysis can be explained by our understanding of the scientific theory underlying the phenomenon we are studying. If we have little understanding of the phenomenon, then our expectations may be quite diffuse or far from the truth and there might be little that surprises us in the data. If our understanding is thought to be quite good, then our expectations might be more narrow and perhaps closer to the truth. 

How can we explain the difference between the data analytic output and our expectations for what we should observe? We would argue that answering this question is a fundamental task for the data scientist. It presents itself in almost every data analysis, large or small, and appears repeatedly. Given that we generally do not observe the truth, the output and the expected outcome are all we have to work with, and juggling these two quantities in our minds is a constant challenge. If the data analytic output is close to our expectations, so that the output is \textit{as-expected}, we might be inclined to move on to the next step in our analysis. If the output is far from our expectation, so that the output is \textit{unexpected}, then we might be inclined to pause and investigate the cause of this unexpected outcome.

\subsubsection{Example: Estimating Pollution Levels in a City}
\label{sec:examplepollution}

Consider a study designed to estimate ambient particulate matter air pollution levels in a city. A simple initial approach might involve deploying a single pollution sensor outdoors that collects hourly measurements of pollution. In order to estimate the daily average level, we could take measurements from a single day's sampling period and average over the 24 hourly measurements. We know that in the United States, the national ambient air quality standard for daily average fine particulate matter is 35~$\mu$g/m$^3$. Therefore, we might expect that the measurement of the daily average that we take should be less than 35~$\mu$g/m$^3$.

Now suppose we take our measurement and discover that the daily average is 65~$\mu$g/m$^3$, which is far higher than what we would expect. In this situation, we might be highly motivated to explain why this deviation has occurred and there may be several possible explanations: (1) Perhaps our interpretation of the ambient air quality standards was wrong and values over 35~$\mu$g/m$^3$ are permissible; (2) there may be a bias in the sensor that causes values to generally read high; or (3) we may have removed measurements below the detection limit before computing the mean, biasing our calculation upwards. Each of these possible explanations represents an error in the way we think about our expectation, our measurement technology, or our statistical computation, respectively.

We could also consider the possibility that we observe a daily average that is 20~$\mu$g/m$^3$. Such an observation would be well within our expectation, and we might be inclined to do nothing in response, but should we? With results that are as-expected, it might be wise to consider what if something had in fact gone wrong or failed and ask how would that occur? Perhaps the sensor was placed in a manner that restricted airflow, reducing the number of particles it could detect. Or perhaps our software for reading in the data erroneously eliminated large values as ``implausible".

In this example, regardless of whether the results are as-expected or unexpected, we need to reconcile what we observe with our understanding of the systems that produced the data and our expectations. Problems may lie in our knowledge of the domain, the technology we deploy to collect data, and the analytic tools that we use to produce results.

\subsubsection{Reconciling Unexpected and As-Expected Results}

Tukey noted the distinction between as-expected and unexpected results and further commented that deciding where to draw the line between the two was a matter of judgment. In a section describing the choice of a cutoff value for identifying ``wild shots" or ``spotty data", he writes
\begin{quote}
The choice [of cutoff] is going to be a matter of judgment. And will ever remain so, as far as \textit{statistical theory} and \textit{probability calculations} go. For it is not selected to deal with uncertainties due to sampling. Instead it is to be used to classify situations in the real world in terms of how it is ``a good risk" for us to \textit{think} about them. It is quite conceivable that empirical study of many actual situations could help us to choose...but we must remember that the best [choice] would be different in different real worlds.
\end{quote}
Determining what is expected from any data analysis and what is unexpected will generally be a matter of judgment, which will change and evolve over time as experience is gained. Tukey ultimately classifies the data points in his case study into three categories as ``in need of special attention", ``in need of some special attention", and ``undistinguished"~\citep{tuke:1962}.

Reconciling the observed output with the expected outcome is an aspect of what Grolemund and Wickham call a sense-making process where we update our schema for understanding the world based on observed data~\citep{grol:wick:2014}. However, rather than take the data for granted and blindly update our understanding of the world based on what we observe, a key part of the data scientist's job is to investigate the cause of any deviations from our expectations and to provide an explanation for what was observed. Should the output be as-expected, it is equally important for a data scientist to consider what might have gone wrong in order to identify any faulty assumptions or logic. With either unexpected or as-expected output, the data scientist must interrogate the \textit{systems} that generated both the output and our expectations in  order to provide useful explanations for the observed results of the analysis.

\subsection{A Systems Approach}
\label{sec:systems}

When confronted with a result that is either as-expected or unexpected, the data scientist's first task is to ask the question, ``How did we get to this point?" Given that we are comparing an observed data analytic output to an expected outcome or range of outcomes, there are two possible avenues to investigate. The first has roots in the \textit{system} that generated the data analytic output and the second has roots in the \textit{system} that generated our expectations for the outcome. There is also a third system to investigate which is the specific software implementation of a data analytic system that was used to generate the output. Thinking of data analysis as, in part, the construction and development of systems is a useful generalization that has some interesting downstream implications. This framing provides a rationale for applying design thinking principles to data analysis development~(Section~\ref{sec:design}) and offers a formal framework for interpreting results that are unexpected.

\subsubsection{Data Analysis Systems Development}

The data analytic system is typically one that is most under the control of the data scientist. This system consists of a series of connected components, methods, tools, or algorithms that together generate the data analytic output. The system may have various sub-systems that themselves branch off into methods, tools, or algorithms. For example, a simple ``data cleaning" sub-system might involve reading in a comma-separated value (CSV) file, removing rows that contain NA values, converting text entries into numeric values, and then returning the cleaned dataset. The input to this sub-system is a CSV file and the output is the cleaned dataset. A second ``summary statistics" sub-system might  take the cleaned dataset and return a table containing the mean and standard deviation of each column.

A system that operates in parallel with the data analytic system is what we refer to as the ``scientific system". This system is built by those knowledgeable of the underlying science and summarizes prior work or preliminary data relevant to the question at hand. The product of the scientific system is some summary of evidence or background information that can be used in conjunction with the design of the data analytic system to predict the output of the data analytic system. The scientific system can be built by the analyst if the analyst is knowledgeable in the area. Otherwise, a collaboration can be developed to build a scientific system whose output can be given to the analyst.

The third system is the software implementation of the data analytic system. Here the analyst chooses what specific software toolchains, programming languages, and other computational environments will be used to produce the output. Some components may be written by others (and perhaps accessed via application programmer interfaces) while others may need to be written by the analyst from scratch. The development of the software system may be dictated by the work environment of the analyst. For example, an organization may have previously decided to only use an existing programming language, toolchain or workflow, limiting the options available to the analyst~\citep{park:2017}.

One important issue that we have left out is the data generation process, which has its own complex systems associated with it. Experimental design, quality assurance, and many other data collection processes will affect data quality and could subsequently cause problems, unexpected outcomes, or failures in the data analysis process. The data scientist will likely have some knowledge of this process and familiarity with the data generation will aid significantly in interpreting the results of an analysis. However, the data scientist may not have much direct control over the data generation and therefore it may be more important to develop strong collaborations with others who may manage the development of those systems. Any discussion of systems must draw useful boundaries around the system and so we have chosen the exclude the data generation process from the discussion of data analysis. We will return to this topic later when discussing the disagnosis of unexpected outcomes and the broader context in which a data analysis sits~(Section~\ref{sec:ethics}).

\subsubsection{Data Analytic Iteration}

These three systems---data analytic, scientific, and software---are the responsibility of the data scientist, regardless of whether the data scientist is entirely responsible for building them. Fundamentally, the data scientist must understand the behavior of each of these systems under typical conditions. In particular, considering how these three systems will interact \textit{before} running the actual data through them is an important aspect of the concept of pre-registration~\citep{asendorpf2013recommendations}. Developing the range of expected outcomes reflects our current state of knowledge and informs our interpretation of the eventual data analytic output. Ultimately, the data analytic output gives us information about each of these systems and how they are operating.

The need for having such a deep understanding of each of these three systems becomes clear when considering the possible outcomes of an analysis. Unexpected outcomes may need to be investigated and their root cause identified. Because the cause of unexpected outcomes can be traced back to any component of any system, it is essential that the data scientist have some knowledge of these systems or collaborate closely with someone who does. For example, a data scientist who is unfamiliar with how air pollution data are collected may be inclined to think that an unexpectedly high value is caused by a software error or a specific algorithmic choice simply because that is the area of greatest familiarity. But unexpected results can just as likely be caused by a misunderstanding of the underlying science or previous work as they can be caused by an error in the software implementation of a statistical algorithm. Results that are as-expected do not escape scrutiny simply because they do not fall outside the range of expected outcomes. Rather, it can be critical to determine if there are faulty assumptions built into the systems that have generated the as-expected output.

The process of (1)~building data analytic, scientific, and software systems, (2)~examining their outputs and comparing them to expectations, (3)~re-examining the design of the system and its relationship with the observed output (whether unexpected or as-expected), and (4)~possibly modifying the systems as a result of observing the results describes the basic \textit{data analytic iteration}. In practice it is not realistic to assume that complex systems can be built in a single pass. Typically, there will be significant uncertainties about the science, the data, the behavior of the methods, and the software, and so substantial testing and iteration will need to be employed (perhaps via simulation). As the systems are tested, and we learn more about the data and how the methods and software behave, we can move forward and refine the systems. For example, in the early stages we might implement more exploratory types of methods in order to learn about the data generation process. In the later stages, as confidence in our knowledge grows, we may implement more efficient procedures that maximize the information obtained from the data.

\subsubsection{Example: Data Cleaning}

\begin{figure}[tbh]
\centering
\includegraphics[width=5in]{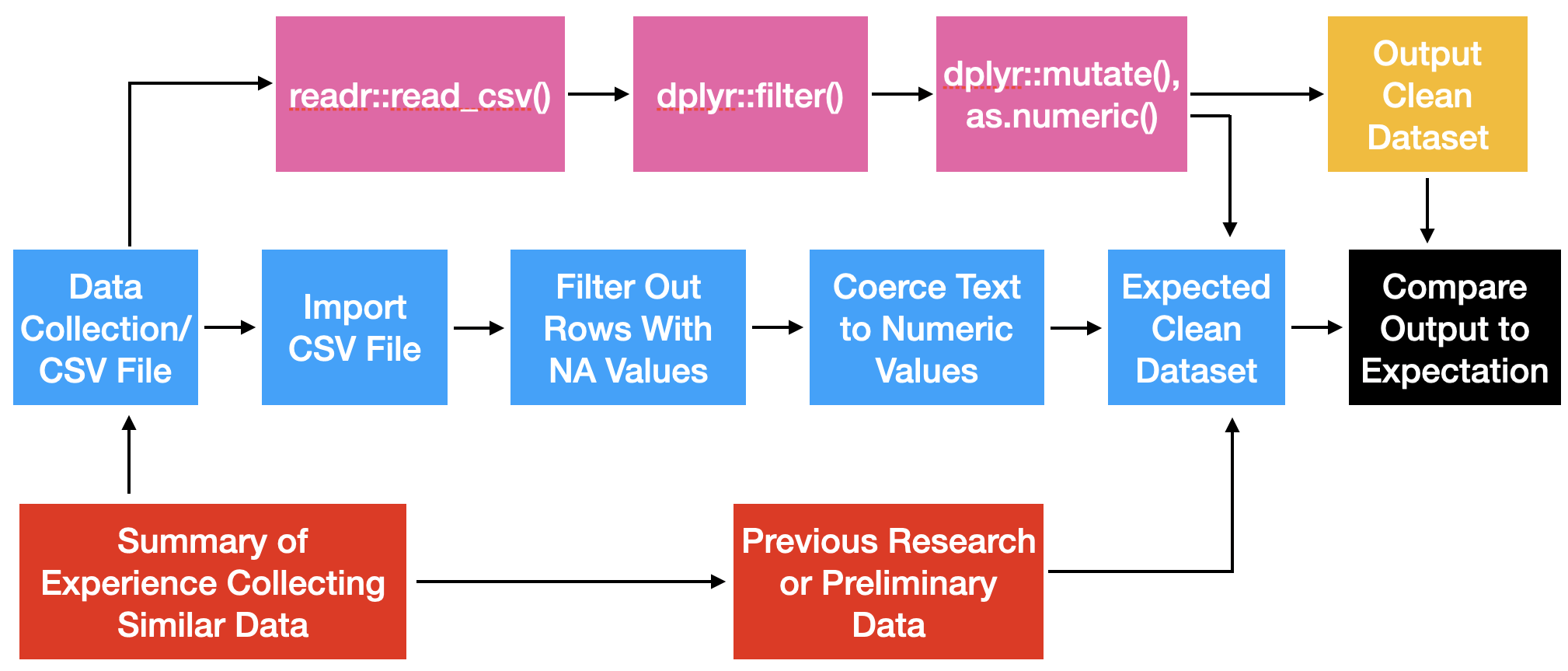}
\caption{Hypothetical data cleaning task with data analytic (blue), scientific (red), and software (pink) systems.}
\label{fig:datacleaning}
\end{figure}

A common ``data checking" task might be to first take the cleaned dataset from the ``data cleaning" sub-system and count the number of rows in the dataset. Continuing the example from Section~\ref{sec:examplepollution}, we might be importing data produced by a remote pollution sensor on a monthly basis in order to monitor environmental conditions. Such data might arrive in the form of a comma-separated-value (CSV) file. Figure~\ref{fig:datacleaning} shows a sample diagram of how the data cleaning sub-system might be organized and implemented. 

If the analyst knew in advance roughly how many rows were in the original raw dataset (say $100$), then the expectation for the number of rows in the cleaned dataset might be something close to $90$ (i.e. 10\% of rows contained NA values). If the result of counting the rows was $85$, then that might be considered ``close enough" and warrant moving on. However, if the number of rows was $10$ or even zero, then that result would be unexpected.

How does an analyst come up with an expectation that approximately 10\% of rows will contain NA values? From Figure~\ref{fig:datacleaning} we can see that the "Expected Clean Dataset" is informed by all three systems---data analytic, scientific, and software---and our understand of how they operate. One possible path is through knowledge of the sensor mechanism, which might be known to be unreliable at times leading to about 10\% of observations having some sort of problem. Another way to develop an expectation is to have knowledge about the underlying scientific problem, which might involve very difficult measurements to take and 10\% missing data is standard for the field. Finally (and perhaps least likely), it may be that the data collection process is fine, but it is known that the software implementation of the data cleaning sub-system is unreliable and occasionally corrupts about 10\% of observations. 

Now suppose that the observed result is that the cleaned dataset has 10 rows, which is unexpected. The analyst can track down the possible causes of this unexpected result by tracing back through the three systems to see if there might be some problem with or misunderstanding about how any of these systems operate. The analyst must evaluate which of these systems is most likely at fault, pinpoint the root cause if possible, and implement any corrective action if the outcome is undesirable. Perhaps an extra step should be added to the data cleaning sub-system that produces an error message if the cleaned dataset has fewer than a certain number of rows. Or perhaps a call should be made to the data collection team to see if there were any recent problems in the latest batch of data. If so, then a protocol can be put in place where the data collection team messages the data scientist if a future batch of data has greater than expected NA observations.

The purpose of laboring through this hypothetical example is (1)~to demonstrate the complexity and variety of knowledge that may be required in order to execute even a simple data checking step, and (2)~to indicate that source of an unexpected outcome can lie in systems beyond the implementation in software. Using more complex systems, like statistical modeling or inference systems, will require further knowledge of data analytic systems, science, and software.

\subsection{Data Analytic Design}
\label{sec:design}

The previous sections have attempted to describe in a general way the activities of a data scientist while doing data analysis. However, a key missing element is the framework for determining how the multitude of data analytic choices and conflicts should be resolved. Traditional descriptions of data science or of data analysis often do not explicitly acknowledge the presence of conflict. Rather, choices are often determined by maximizing or minimizing some arbitrary quantitative criterion~\citep{tuke:1962}. However, conflicts can arise from many different aspects of data science work, many of which originate outside of the data or the data analysis. As a result, many conflicts cannot be resolved explicitly by employing the tools of data analysis, but nevertheless can affect the conduct of a data analysis or even cause the analysis to fail~\citep{robi:noli:2020,mcgowan2021design}. For better or worse, resolving conflicts is commonly part of the data scientist's job.

It is useful to state explicitly that a data analysis is a technical product that must be created and brought into existence. Without the presence of a data scientist, a data analysis would not naturally occur. The design and production of technical systems is by no means a new concept or field of study~\citep{brooks1995mythical,nasasystems2007,Cross2021divergent,Cross2011design}. However, considering a data analysis as a technical product to be designed, produced, maintained, and ultimately retired, is not a common perspective in the various data science-related sub-fields, including Statistics~\citep{park:2017}. 

The need for design thinking in data analysis is driven by the presence of conflicts introduced by various constraints on the analysis development. Constraints imposed by budget, time, expertise, personnel, the audience, or other stakeholders, can fundamentally change how the data analytic, scientific, and software systems are built and what the output looks like at the end. Indeed, without such contraints, there is generally little need for design thinking. Analyses that are done on a very short timescale can look different from otherwise similar analyses done on a longer timescale. Analyses developed with large budgets will look different from analyses done on a shoestring. Computational resources often affect what types of methods can be executed within the available time. Analyses presented to the CEO will need to look different from analyses presented to the principal data scientist. The reality of data analysis development is that data scientists must produce the best product within the numerous constraints imposed from the outside (and this is before we consider the data themselves). If the constraints make developing an analysis untenable, then the data scientist may need to negotiate with a stakeholder to make some changes~\citep{robi:noli:2020}. 

\subsubsection{Design Principles}

Limits placed by stakeholders and contextual factors are not the only considerations for the data scientist, as there is a growing collection of design principles that are being considered to guide the development of data analyses and data analytic products~\citep{park:2017,mcgowan2021design,wood:2019}. These principles serve to characterize a data analysis and to distinguish properties of one analysis from another. As an analysis is developed, the data scientist must choose how much emphasis will be placed on different principles.

 For example, reproducibility is  commonly cited as an attribute of a data analysis and that most analyses should strive to be reproducible to others~\citep{peng:2011}. However, reproducibility does not always make sense and is not always necessary or possible. Common data products such as dashboards or interactive visualizations often do not produce reproducible analyses because the audience for such products generally does to require it. Quick one-off analyses may not be reproducible if the stakes are very low. Even large-scale analyses may not be reproducible to the general public if the analyses use private or proprietary data~\citep{peng:domi:zege:2006}. 

The extent to which an analysis adheres to certain principles (such as reproducibility) can be driven by numerous outside factors that the data scientist must negotiate before, during, and after the analysis is completed. For example, in the United States, the Health Insurance Portability and Accountability Act (HIPAA), which became effective around the year 2000, greatly limited the reproducibility of data analyses using personal health data. Since 2000, data scientists using identifiable health data in analyses must sacrifice some reproducibility or else find a way to anonymize the data. On the other hand, many journals now require that data be deposited in third-party repositories so that analyses have a chance at being reproduced if needed~\citep{paltoo2014data}. Requirements regarding reproducibility and privacy will likely come into conflict and may alter the nature of a data analysis in order to balance certain tradeoffs. For example, analyses of aggregated data may be less desirable in some cases because of ecological bias~\citep{wake:shad:2006}, but may be more reproducible because privacy concerns diminish with increasing aggregation.

Another example of a design principle to consider when building a data product is the level of skepticism presented~\citep{mcgowan2021design}. Skepticism can be characterized as the exploration of multiple alternative hypotheses that may be consistent with the observed data. While healthy skepticism might be considered a bedrock element of scientific inquiry, it can be disorienting and distracting in some circumstances and with some audiences. Machine learning algorithms implemented to predict what web site users might want to buy do not present any skepticism, nor is any likely to be welcome. However, during the development of the underlying algorithm, some skepticism might be useful when discussing the algorithm with other analysts or engineers. Here, the audience for the analysis  or analytic product plays a significant role in shaping the analysis and how it is presented.

There may be other design principles that are valuable in guiding the development of a data analysis and the community of data scientists will have to formalize them as the field develops. Most likely, these principles will evolve over time as technologies, methodologies, culture, and the data science community continue to change. For example, reproducibility did not receive much attention until computing and the Internet became fundamental aspects of data analysis and academic research~\citep{schw:karr:clae:2000}. New technologies like git allow for analyses to be version controlled and create more collaborative opportunities~\citep{park:2017}. More generally, cloud-based platforms allow for collaborative data sharing~(Figshare, Open Science Framework), paper writing~(Overleaf, Google Docs, arXiv), and coding~(GitHub, GitLab).  Similar to other areas where design is an important consideration, the principles that guide the development of products must keep up with the standards of the times~\citep{Cross2011design}.


\subsection{Data Science Ethics}
\label{sec:ethics}

Data science applications are investigations of the world and the very act of conducting such investigations can have an impact on the world, intended or not. Therefore, it is critical to consider and discuss what those impacts may be and whether the potential benefit is worth the risk~\citep{loukides2018ethics}. Traditional scientific investigations that collect new data (especially from humans) generally have to be reviewed by an institutional review board in order to ensure that the methods adhere to scientific standards and that the tradeoff between risk and benefit is properly balanced. While data science applications often feel different from traditional scientific studies in that the data come from a different type of source (e.g. scraped off the public web or pulled from a database), the similar basic principles should be considered~\citep{goodyear2007declaration,boyd2012critical}. Even without any explicit data collection, a data scientist tasked with developing an algorithm has ethical considerations to make, for example with regard to bias, fairness, and accountability~\citep{danks2017algorithmic,rosenblat2014algorithmic,leonelli2021fair}. Recently, there has also been discussion of data science oaths, similar to the Hippocratic oath taken by doctors, that would give data scientists explicit priniciples to which they would pledge to adhere~\citep{national2018data,loukides2018ethics}.

Data analyses must be considered within a specific surrounding context and knowledge about that context can change the way an analysis is conducted, if at all. In their book \textit{Data Feminism},  Catherine D'Ignazio and Lauren Klein write that ``a feminist approach [to thinking about datasets] insists on connecting data back to the context in which they were produced. This context allows us, as data scientists, to better understand any functional limitations of the data and any associated ethical obligations, as well as how the power and privilege that contributed to their making may be obscuring the truth."~\citep{dign:klei:2020} Data analyses that accept the data as given ignore this obligation to the context from which the data were generated and to which the results will be presented. The risk of ignoring this context is producing an analysis that is incorrect or not useful, at best, and unethical or harmful, at worst.

\section{Core Data Science Tooling}

The area of data science that is best developed is the area of software tooling. Numerous software packages and systems have been developed for the express purpose of doing data science~\citep[e.g.][]{r,wickham2019welcome,bressert2012scipy,mckinney2011pandas,sas2015base}. These packages are written in a variety of languages and implemented on many platforms. The existence of such a diversity of tools might lead one to conclude that there is little left to develop. Of course, the vibrant and active developer communities organized around both the R and Python programming languages, to name just two, is evidence that new tools need to be continuously developed to solve new problems and handle new situations.

It is worth singling out the R programming language here in order to highlight its historical origins with the S language. S was developed at Bell Labs to address an important and novel problem: a language was needed to do interactive data analysis. In particular, exploratory data analysis was a new concept that was difficult to execute in existing programming systems. Rick Becker, one of the creators of the S language, writes in ``A Brief History of S,"
\begin{quote}
[W]e wanted to be able to interact with our data, using Exploratory Data Analysis (Tukey, 1971) techniques.... On the other hand, we did occasionally do simple computations. For example, suppose we wanted to carry out a linear regression given 20 x,y data points. The idea of writing a Fortran program that called library routines for something like this was unappealing. While the actual regression was done in a single subroutine call, the program had to do its own input and output, and time spent in I/O often dominated the actual computations. \textit{Even more importantly, the effort expended on programming was out of proportion to the size of the problem}. An interactive facility could make such work much easier. [emphasis added]
\end{quote}
The designers of the S language wanted to build a language that differed from programming languages  at the time (i.e. Fortran), a language that would be \textit{designed for data analysis}. It is worth re-visiting the idea that tools could be designed with data science problems specifically in mind. In particular, it is worth considering what tools might look like if they were designed first to deal with data analysis rather than to write more general purpose software.

A simple comparison can provide a demonstration of what we mean here. Consider both the Fortran and R languages. Fortran, like many programming languages, is a compiled language where code is written in text files, compiled into an executable, and then run on the computer. R is a an interpreted language where each expression is executed as it is typed into the computer. R inherits S's emphasis on interactive data analysis and the need for quick feedback in order to explore the data. Designing R to be an interpreted language as opposed to a compiled language was a choice driven by the intended use of the language for data analysis~\citep{ihak:gent:1996}.

\subsection{Data Analysis Representation}

The presentation of a data analysis may come in a variety of forms---a report, a slide deck, a journal publication, or even a verbal presentation. Beyond this final presentation form, there is an expectation today that a data analysis can be communicated in a different form with greater details of how the results were produced. But a question arises as to what is the appropriate manner in which a data analysis should be represented. What is the best way to represent the ``source code" of a data analysis?

Before we can attempt to answer this question we need ask what is the purpose of having any representation of a data analysis other than the final outputs. There are a few reasons that come to mind. First is reproducibility. The importance of reproducibility, as noted above, can vary with the context but is absolutely critical in a scientific setting. Providing a detailed representation of an analysis allows independent data scientists to reproduce the findings, which in general may be valuable in order to build trust.

Reproducibility alone is often not valuable in and of itself---if a result is reproduced by executing the same code on the same dataset, then essentially we have not learned much as our expectation was that the result should reproduce. However, if a result does \textit{not} reproduce, then having some detailed representation of the analysis is critical. This brings us to a second purpose for data analysis representation, which is diagnosing the source of problems or unexpected findings in the analysis. 

A third reason for having access to the details of an analysis is to be able to build on the analysis and to develop extensions. Extensions may come in the form of sensitivity analysis, or the application of alternate methods to the same data. A detailed representation would prevent others from having to re-do the entire analysis from scratch.

The current standard for providing the details of a data analysis is providing the literal code that executed the steps of an analysis. This representation works in the sense that it typically achieves reproducibility, it allows us to build on the analysis, and it allows us to identify the source of potential unexpected results, to some extent. However, the computer code of an analysis is arguably incomplete, given that a complete data analysis can be composed of other systems of thinking, such as the data analytic and scientific systems described in Section~\ref{sec:systems}. The R code of a data analysis generally does not have any trace of these systems. One consequence of this incompleteness is that when an unexpected result emerges from a data analysis, the code alone is insufficient for diagnosing the cause. Literate programming techniques provide a partial solution to this issue by providing tooling for mixing computer code with documentation~\citep{knut:1984}.

Data scientists can perhaps learn lessons from designers of traditional programming languages. Over time, as computers have become more powerful and compilation technologies have advanced, programming languages have placed greater emphasis on being relevant and understandable to human beings. A purpose of high-level languages is to allow humans to \textit{reason} about a program and to gain some understanding of what critical activites are occurring at any given time. Data scientists may benefit from considering to what extent current data science tools and approaches to representing data analysis allow us to better \textit{reason about an analysis} and potentially diagnose any problems in design.

\subsection{Data Analytic Quality and Reproducibility}

Recent work has focused on the quality and variability of data analyses published in various fields of study~\citep[e.g.][]{open2015estimating,jager2014estimate,ioannidis2005most,patil2016should}. Standards on computational reproducibility have improved the transparency of analyses over the past 20 years and have arguably allowed problems to be identified more quickly~\citep{peng2021reproducible}. For example, if an independent data scientist knows that an analysis is reproducible, then more time and resources would be available to understand the higher order aspects of the analysis, such as why a given model was applied, instead of wasting time on figuring out what version of software was used. Given that some pathological examples of irreproducibility have taken upwards of 2,500 person-hours to diagnose, reducing the time to identify the root causes of problematic analyses is certainly welcome~\citep{baggerly2009deriving,baggerly2010disclose,goldberg2014duke}.

While it might be argued that reproducible analyses meet a minimum standard of quality~\citep{peng:domi:zege:2006}, such a standard is insufficient if only because reproducible analyses can still be incorrect or at least poor quality~\citep{leek:peng:2015}. Reproducibility is valuable for diagnosing a problem after it occurs, but what can be done to prevent problems before the are published or disseminated widely? What tools can be built to improve the quality of data analysis at the source, when the analysis is still in the data scientist's hands?

The systems approach described in Section~\ref{sec:systems} naturally leads one to consider the concept of explicitly testing expectations versus outputs. Here, we can potentially borrow ideas from software testing that might help to improve the conduct of data analysis more generally~\citep{tdda:2016,testthat2011}. There may be other ideas that can be adapted for the purpose of improving data analysis and data scientists should continue to work to build such tools.

\subsection{Diagnosing Data Analytic Anomalies}

The complexity of systems that data scientists must build to analyze data makes diagnosing problems or even unexpected results a challenge. Characterizing anomalies in a data analysis requires that we have a thorough understanding of the entire system of statistical methods that are being applied to a given dataset in addition to the scientific background and the software implementation. The data analytic system alone will include traditional statistical tools such as linear regression or hypothesis tests, as well as methods for reading data from their raw source, pre-processing, feature extraction, post-processing, and any final visualization tools that are applied to the model output. Ultimately, anomalies can be caused by any component of any system, not just the data analytic system, and it is the data scientist's job to understand the behavior of the entire system. Yet, there is little direct tooling that considers the complexity of such systems and how we might diagnose problems that occur within them.

Shifting to a framework that focuses on the design and implementation of systems for data analysis opens up new avenues for thinking about the data analysis process more generally and how we might represent that process in a useful way. In particular, we can apply approaches from systems engineering to formally model how data scientists think about a problem. Tools like fault trees may provide a way of evaluating how well analysts understand the complex systems being applied to their data and can suggest improvements to those systems~\citep{vesely1981fault}. Data analyses are often idiosyncratic and highly specific to the scientific question and data. To the extent that we can develop tooling to help us better characterize the data analytic process, we will be able to improve the quality of data analysis and detect problems before they can cause any damage.


\section{Building a Theory of Data Science}

In Section~\ref{sec:design} we discussed the importance of design thinking in data analysis development and concluded that design principles for data analysis likely would evolve naturally over time. The idea of an ever-changing landscape in which data analysis is to be conducted may produce discomfort in some statisticians or data scientists who conceive of data analysis as a task driven by universal characteristics or quantities. However, it is already clear that technology alone can transform the practice of data analysis and changes to standards or cultural norms can make some data analyses infeasible or unethical.

Given this setting, what hope is there in developing any sort of theory of data science? One answer is that any theory that might be developed would likely look very different from the kind of theory with which statisticians are familiar. It is unlikely that we will discover many universal truths in the pursuit of data science theory, given the inherently changing nature of the field itself. However, we can draw lessons from experiences in other similar fields as well as recent experience within the field of data science itself.

One perspective on the traditional theory of statistics is that we can summarize our past experience using various data analytic tools in order to make universal claims about the future, under various important assumptions, of course. A theory of data science will likely share the first part of that perspective (summarizing the past) but lack the second part (making universal future claims). Nevertheless, there is value in summarizing past experience, as we can see from observing other fields of study.

\subsection{Summarizing Experience}

Artistic fields all have their own theories that serve to summarize past experience. For example, music theory is largely a descriptive practice that attempts to draw out underlying patterns that appear in various forms and instances of music~\citep{aldwell2018harmony}. Music theory has produced concepts like ``Sonata Form", which says that a piece of music has an exposition, a development, and a recapitulation. We also have tonal harmonic analysis, which provides a language for describing the harmonic transitions in a piece of music. We can easily find chorales written by Johann Sebastian Bach that follow similar harmonic patterns as songs written by contemporary composers. In this way,  tonal theory allows us to draw connections between very disparate pieces of music. One thing that the tonal theory of harmony does \textit{not} give us is a recipe for what to do when creating new music. Simply knowing the theory does not make one a great composer. 

Arnold Schoenberg, in his textbook on the theory of harmony, \textit{Harmonielehre}, argued strongly against the idea that there were certain forms of music that inherently ``sounded good" versus those that ``sounded bad"~\citep{schoenberg1983theory}. He argued that the theory of music tells us not what sounds good versus bad but rather tells us what is \textit{commonly done} versus \textit{not commonly done}. In other words, it is a summary of experience. One might infer that things that are commonly done are therefore good, but that would be an individual judgment and not an inherent aspect of the theory. As Schoenberg is perhaps best known as a leading \textit{atonal} composer, it would seem that having a summary of what was commonly done in the past served as a valuable guide to what \textit{not} to do in the future. Such an outcome is in fact in line with what Tukey would have recommended, which is that theory should ``guide, not command."~\citep{tuke:1962}

Established experts in the field of data science are likely to think about data science activities differently from novices. Experts understand and recognize how to approach different problems, how to navigate various contextual issues, and can usually make informed judgments about what tools to apply at any given moment. It is easy to forget that not all data scientists are experts yet. Novices or newcomers to the field will need some information to start with, to get them going on a new project. A data scientist approaching a new problem involving an environmental health question could benefit from having a concise summary of past experience analyzing air pollution and health data. A new data scientist confronted with evaluating a new web site layout might benefit from a summary of past experience conducting A/B tests on a web platform. In either case, the data scientist may find that the previous experiences do not exactly fit their current situation, but there may be ways to adapt previous approaches to the new problem. 

One useful example considers the vast and nebulous work of data cleaning. Transforming so-called raw data into usable tidy data is arguably a task that is truly specific to the problem and data at hand. Yet, there may be commonalities in data cleaning that span various applications. For example, Broman and Woo discuss their experience working with spreadsheet data, particularly in the context of collaborating with non-statisticians~\citep{broman2018data}. Although some may consider spreadsheets a poor tool for data storage and representation, it is the ideal (or perhaps the only) tool for others. Broman and Woo offer reasonable guidelines for using spreadsheets while preserving the integrity of the data and facilitating future data analyses.

\subsection{Abstracting Common Practice}

A lesson we can draw from recent experience in the data science field comes from software. Software plays an important practical role in allowing data scientists to actually analyze data. But perhaps paradoxically, it also plays an important theoretical role in summarizing and abstracting common data analytic routines and practices. A prime example comes from the recent development of the ``tidyverse" collection of tools for the R programming language~\citep{wickham2019welcome}, and in particular the dplyr package with its associated verbs for managing data. This collection of software packages has revolutionized the practice of data analysis in R by designing a set of tools oriented around the theoretical framework of tidy data~\citep{wickham2014tidy}. This framework turns out to be useful for many people by abstracting and generalizing a common way of thinking about data analysis via the manipulation of data frames. It is perhaps one of the most valuable theoretical constructs in data science today.

Yet, is it true that all data are tidy? No, reasonable counterexamples abound in every scientific context. Is it even true that all data analyses can be accomplished within the tidy data framework? No, for example, there are some analyses that are better suited to using matrices or arrays. Will the tidyverse continue to be useful forever into the future? Probably not, because new tools and frameworks will likely be developed. Indeed, there are unlikely to be any universal truths that emerge from the tidyverse software collection. But it is nevertheless undeniable that the tidyverse has provided a useful structure for reasoning about and executing data analyses. 

A key purpose of writing software is to codify and automate common practices. The concept of "don't repeat yourself" encapsulates the idea that in general, computers should be delegated the role of executing repetitive tasks~\citep{thomas2019pragmatic}. Identifying those repetitive task requires careful study of a problem both in one's own practice and in others'. As we aim to generalize and automate common tasks, we must simultaneously consider whether the task itself is unethical or likely to cause harm. With the scale and speed of computing available today, locking in and automating a process that is biased can cause significant harm and may be difficult to unravel in the future~\citep{oneil2016weapons}.

\subsection{Lessons Learned}

Developing theoretical abstractions or generalizations requires that we look across the field of data science and identify commonalities, whether they are methodological, practical, or otherwise. Such commonalities can be formalized for the purposes of communicating them more broadly to the community of data scientists and to provide a perspective on current practices. Such generalizations should not be thought of as absolute \textit{laws} of data science, but rather useful summaries or perhaps guidelines. If anything, the theory can serve as a fallback for individual data scientists in situations where there is little other information on which to act.

Successful data science projects face a difficult choice regarding what to present and what lessons to draw. Typically, the obvious result to present is the answer to the scientific, business, or policy question that originally initiated the data science effort. In an academic setting such results may be published in a scientific-area journal. There may be second order results to present regarding any methodologies or techniques that have been developed as part of the project that can separated out as an independent component. Results about new techniques might be published in a methodological journal. If there is any knowledge gained about the process of doing data science, or data analysis more specifically, there is no obvious venue in which to publish such information.

If data analysis is an important part of data science, then there needs to be a place to communicate to other members of the field any lessons learned from doing data analysis. The existing publication process is insufficient for this purpose because descriptions of data analyses in journal articles are primarily focused on making the work reproducible. As a result, typically the minimum amount of information is presented in order to generate the published results. Currently, communications and discussions about the process of doing data analysis occur on blogs, social media, and various other informal channels. While such accounts are often useful, there could be a benefit to creating a more formal approach to documenting, summarizing and communicating lessons learned from data analysis~\citep{waller2018documenting}. Such an effort would give novices an obvious place to learn about data analysis and would give researchers in the field a way to see across the experiences of other data scientists in order to identify any common structures. If we are to treat data science as any other scientific field, we need a common venue in which we can discuss lessons learned and identify areas that we do not yet understand or where there are knowledge gaps.

\section{Teaching Data Science}

The teaching of statistics and now data science has evolved substantially over the past three decades with an increasing recognition of the unique nature of statistical thinking~\citep{wildpfannkuch1999,grolemund2014cognitive,lovett2000applying}. The American Statistical Association's Curriculum Guidelines for Undergraduate Programs in Statistical Science~\citep{asaundergrad2014} and the Guidelines for Assessment and Instruction in Statistics Education~\citep{franklin2007guidelines} both emphasize the need for real problems and datasets to demonstrate the messy nature of data analysis. More recent guidelines for undergraduate data science programs have highlighted integration with the sciences and the importance of algorithmic thinking and software development~\citep{de2017curriculum}.  Other recommendations have similarly emphasized computing and the idea of ``thinking with data"~\citep{nolanlang2010,hardin2015}. In general, we have moved away from thinking about statistics and data science education as an assortment of tools and methods towards a more integrative approach that focuses on the scientific method and its relation to statistical analysis~\citep{asaundergrad2014,de2017curriculum}.

The teaching of data science is hampered by the simple fact that no tool or method is used in every analysis and no analysis requires every tool. As a result, the teaching of data science can at times feel highly inefficient and perhaps \textit{ad hoc}. One often resorts to teaching a handful of tools and presenting a handful of case studies and then drawing a loose graph connecting the two sets together. If one does not present a sufficently diverse set of case studies, the students may not have the opportunity to apply  all the tools. Similarly, if one does not teach a complete set of tools, students may not be able to address every case study. One solution to this matching problem is to force students to apply specific tools to specific problems, but this scenario generally will not reflect how data science works in the real world. 

The matching of tools to case studies can work in the sense that students often walk away with a useful education. However, this approach can sometime lead students to believe that for every case study there is a ``correct" set of tools to use and for every tool, there is a ``correct" set of applications. In reality, many applications will admit a variety of tools that can lead to the same basic conclusion and the choice of tooling will often be determined by factors outside the data. There is unfortunately no way around the conundrum of multiple tools being applicable for a given case study and given the limits on time and resources for most teachers, cases will usually be presented using a single approach. However, even if time were available to present multiple different approaches to solve a problem, students might then be inclined to ask, "Which is the best way?"

In our experience teaching data science in the classroom at the graduate level, the structure of a data science course is often best defined by what it is \textit{not}. Typically, it is not whatever is taught in the rest of the core statistics curriculum~\citep{kross2020democratization}. While this approach can sometimes result in a reasonable course syllabus, it is hardly a principled approach to defining the core curriculum of an area of study. On the other hand, many guidelines for building data science programs are indistinguishable from statistics programs~\cite{de2017curriculum}. Until we have a clear vision for what consists of the core of data science, there will likely not be any better alternative proposals for building a coherent data science program. Moving forward, the danger of not having a focused vision is that data science education will devolve into teaching a never-ending proliferation of topics related to data.

An apprenticeship model is sometimes proposed as an ideal way to teach data science, particularly in non-degree ``bootcamp" style training programs~\citep{craig:2020,lambdaschool:2021,gassembly:2021,ibm:2020}. Such one-on-one attention with a mentor or advisor is indeed valuable, but is arguably the most inefficient approach and is difficult to scale to satisfy all demands for data science education. As a way to teach highly specific skills needed in a particular setting, apprenticeships may be the best approach to training. But as a general model for teaching data science it is likely not feasible. In addition, individualized instruction risks reinforcing the idea that data science is an individualized art and that the student only needs to learn whatever the advisor happens to know. The resulting heterogeneity of education goes against the idea that data science is in fact a unified field with a core set of ideas and knowledge.

Much like with other fields of study, the material that is most suitable for a classroom setting, where all students are given the same information, is theory. Material that is best suited for individualized instruction is specific information required to solve specific problems. If data scientists are able to build a theory, draw generalizable lessons, abstract out common practices, and summarize previous experience across domains, then it would make sense to teach that in the classroom, much like we would teach statistics or biochemistry. The classroom would also be the logical place to indicate what remains unknown about data science and what questions could be answered through further research.

\section{Summary}

Sharpening the boundaries of the field of data science will take considerable time, effort, and thought over the coming years. Continuing on the current path of merging together skills and activities from other fields is valuable in the short term and has the benefit of allowing data scientists to explore the space of what the field could be.  But the sustainability of an independent data science field is doubtful unless we can identify unique characteristics of the field and define what it means to develop new knowledge.

In this review we have made an attempt at describing what are the core ideas about data science that makes data science different from other fields. A key potential target of further exploration is the area of data analysis, which is an activity that continues to lack significant formal structure more than 50 years after Tukey's paper on the future of data analysis. Given the importance of data analysis to the job of the data scientist, developing a more robust formal understanding of the process could provide useful perspective to experienced practitioners and would give teachers of data science a foundation for training newcomers to the field. Furthermore, such a formal basis for data analysis would immediately scale up to meet the significant demands for data science training in academia, industry, and government, much the same way that teaching linear regression as $y =X\beta+\varepsilon$ is more efficient than teaching separate linear regression models for every application.

Like with all fields of study, formalization and theory can only take us so far. Ultimately, in order to achieve concrete outcomes in data science, some level of specialization and individualized knowledge will be required. The frequently-changing nature of tools, technologies, and software platforms may preclude those elements from playing a significant role in any data science theory and practitioners will need to continuously adapt to the latest developments. Striking a balance between the general and the specific is a challenge in any field and will be an important issue that defines the future of data science.

\bibliographystyle{asa}
\bibliography{combined}

\end{document}